\documentclass[aps]{revtex4}

\usepackage{bm}
\usepackage{graphicx}
\usepackage{amsmath}
\usepackage{amssymb}
\usepackage{amsfonts}
\usepackage{float}
\usepackage{color}

\begin{document}

\title{Scaling and universality in glass transition}

\author{Antonio de Candia$^{1,2,3}$}

\author{Annalisa Fierro$^{1,2}$} 

\author{Antonio Coniglio$^2$}

\affiliation{$^1$Dipartimento di Fisica ``Ettore Pancini'', 
Universit\`a di Napoli
  ``Federico II'', \\ Complesso Universitario di Monte Sant'Angelo, via
  Cintia, 80126 Napoli, Italy}

\affiliation{$^2$CNR-SPIN, via Cintia, 80126 Napoli, Italy}

\affiliation{$^3$INFN, Sezione di Napoli, via Cintia, 80126 Napoli, Italy}

\maketitle
\section*{Abstract}
Kinetic facilitated models and the Mode Coupling Theory (MCT) model B 
are within those systems known to exhibit a discontinuous dynamical 
transition with a two step relaxation. We consider a general scaling 
approach, within mean field theory, for such systems by considering the 
behavior of the density correlator $\langle q(t)\rangle$ and the 
dynamical susceptibility $\langle q^2(t)\rangle -\langle q(t)\rangle^2$. 
%a $\beta$ relaxation with
%an approach to a plateau and a departure from the plateau characterized by
%a critical relaxation time $\tau_\beta$, followed by a second relaxation
%time $\tau_\alpha$. 
Focusing on the Fredrickson and Andersen (FA) facilitated spin model on the Bethe 
lattice, we extend a cluster approach that was previously developed for 
continuous glass transitions by Arenzon et al (Phys. Rev. E 90, 020301(R) (2014))
to describe the decay to the plateau, and consider a damage spreading mechanism to describe 
the departure from the plateau. We predict scaling laws, which relate dynamical exponents to the static exponents
of mean field bootstrap percolation. The dynamical behavior and the scaling laws for both density correlator and dynamical susceptibility coincide with those predicted by MCT. These results explain the origin of scaling laws and the universal behavior associated 
with the glass transition in mean field, which is characterized by the divergence of the static length of the bootstrap percolation model with an upper critical dimension $d_c=8$.

\section*{INTRODUCTION}

Many physical systems and models exhibit a sudden slowing down of their dynamics,
followed by a dynamical transition associated with a structural arrest. Roughly, we can distinguish two type of 
transitions, continuous and discontinuous, depending whether or not there is a jump at the threshold of the dynamical 
correlator in the infinite time limit. An example of the first category is given by the sol-gel transition 
(e.g. Refs.~\cite{flo,deg,zac}).
This dynamical transition was studied using a cluster approach, based on percolation theory~\cite{fierro}. 
An explicit scaling form for the dynamical correlator was found, and general scaling laws    
connecting the dynamical exponents with the random percolation exponents were derived.
Recently, it was shown~\cite{arenzon} that using mean field percolation exponents, the same scaling form for the 
correlator and the same scaling relations were also valid for the continuous transition of MCT model A, suggesting 
that the origin of the continuous MCT scaling relations is due to an underlying static transition in the same 
universality class of random percolation. 

%and  suggesting the same cluster mechanism holds 
%more generally for glassy systems with continuous ergodicity breaking \cite{nota_mct}.
The glass transition belongs instead to the second category.  
%Examples of the first category isare  gels (e.g. Refs.~\cite{flo,deg,zac}), and mode coupling theory (model A), 
A great advance in glass theory was provided by MCT developed by 
G\"otze and collaborators \cite{gotze1,gotze2}. This theory starting from first principles, 
under some mean field approximations, predicts a dynamical arrest at a finite 
temperature $T_c$, characterized by power law behavior and universal scaling laws. 
These theoretical predictions have been tested in great detail both experimentally and numerically 
\cite{pusey,kob,dawson,mallamace}.
Other models, like $p$-spin glass models (e.g. Refs.~\cite{pspin1,pspin2}), Random Field Ising model  in an external field (e.g. Refs.~\cite{RFI1,RFI2,RFI3,birolitarjus}), kinetic facilitated models (e.g. Refs.~\cite{KA,FA,sollich}), reproduce in mean field the same dynamical behavior  and scaling laws.
However  the  transition described by MCT does not  seem to exhibit any critical 
change in the structure and no diverging static length.  How can we then  explain the scaling laws and universal behavior found at mean field level? 
In this paper we consider 
as paradigmatic example the Fredrickson and Andersen facilitated Ising 
model~\cite{FA} on a Bethe lattice~\cite{sellitto}. The infinite time limit of 
the persistence of this model, $\Phi(t)$, tends~\cite{sellitto} to the order parameter of the bootstrap percolation (BP) 
model \cite{CLR,schwarz,baxter2011}. The BP model exhibits a mixed order
transition with an order parameter which jumps discontinuously at the transition, nevertheless the fluctuations,  and the critical length associated to it, diverge as the transition is approached from the glassy phase. 
%Using a physical picture for the decay mechanism  
%of the correlator, $\Phi(t)$, based on a combination of a cluster formalism and a damage spreading %approach, 

Generalizing the cluster approach considered for the continuous dynamic transition \cite{arenzon}, 
we are able to predict the dynamical behavior for the correlator and for the dynamical susceptibility of the FA 
facilitated model, including universal scaling laws that relate dynamical exponents with the static universal exponents of BP. Using the mean field values of these static exponents, we find that the dynamical behavior and the scaling laws are the same as predicted by MCT model B, what validates the early suggestion~\cite{sellitto,ArSe,FrSe,maurocrossover} that the 
facilitated model and the MCT model B have a similar dynamical behavior.
Akin the results found for the continuous transition~\cite{arenzon}, using the cluster approach 
we find a new more precise form for the approach of the correlator to
the plateau, characterized by a power law, followed by a stretched exponential divided by a power law. These new predictions are verified  numerically on both FA facilitated model and  MCT model B.
All these results suggest a general common mechanism for discontinuous glass transition at mean field level, based on
a static transition in the same universality class of bootstrap percolation with  
a diverging static length, which is responsible for the origin of scaling  and universality present in such a wide range of  systems, apparently very different from each other.

Here, for convenience, we summarize the main results.
Given a two step relaxation, the correlator can be written as $\Phi(t) -m_c \simeq \epsilon^{\beta}F(t/\tau_\beta,t/\tau_\alpha)$, where  $m_c$ is the value of the plateau at transition, $\beta =1/2$ is the order parameter BP exponent, $\tau_\beta\sim \epsilon^{-z_1}$ corresponds to the first step relaxation to the plateau,  and $\tau_\alpha\sim\epsilon^{-z}$ corresponds to the second step relaxation time. At criticality $\Phi(t)-m_c \sim t^{-a}$,  
with $z_1=\beta /a= 1/2a$, while the approach to the plateau is given by a stretched exponential divided by a power law with precise predictions following from the cluster approach. 
The departure from the plateau is given by $\Phi(t) -m_c \sim -\epsilon^{\beta}(t/\tau_\beta)^b$, which is interpreted  as damage propagating from an initial density of infected sites
$\epsilon^{\beta}$, times  $\left(t/\tau_\beta\right)^b$, the number of distinct  damaged sites  by one initial infected site during the time $t$. 
A consequence of the scaling function of the two variables is the scaling relation between dynamic exponents $a$, $b$, $z$ and the BP static exponent $\beta$,   $z=\beta/a+\beta/b=1/2a+1/2b$.
The dynamical susceptibility, $\chi_4(t) =N (\langle q^2(t)\rangle -­\langle q(t)\rangle^2)$ (where $N$ is the 
number of particles)  in the liquid phase is given by
$\chi_4(t)\simeq\epsilon^{-\gamma}G_{-}(t/\tau_\beta,t/\tau_\alpha)$, where $\gamma =1$ is the BP critical exponent of the fluctuation of the order parameter.
This scaling leads to  $\chi_4(t)\sim t^{a\gamma/\beta}=t^{2a}$ for $t<\tau_\beta$ with a crossover 
to $t^{2b}$ for $\tau_\beta<t\ll\tau_\alpha$. 
This crossover is a consequence that the dynamics in this regime is due to propagation of damage and 
that $\chi_4(t)$ is proportional to the square of distinct damaged sites. 
Finally, $\chi_4(t=\tau_\beta)\sim\epsilon^{-\gamma}=\epsilon^{-1}$ and $\chi_4(t=\tau_\alpha)\sim\epsilon^{-\gamma-2\beta}=\epsilon^{-2}$ and goes to zero in the infinite time limit. In the glassy phase $\chi_4(t)\sim t^{a\gamma/\beta}=t^{2a}$ for $t<\tau_\beta$ with a crossover 
to a constant plateau whose value diverges as $\sim\epsilon^{-\gamma}=\epsilon^{-1}$ .

In the following, using the cluster approach and a damage spreading mechanism, we will derive on the Bethe lattice the dynamical behavior of the correlator and the dynamical susceptibility in terms of critical exponents of the BP model, and  compare with MCT results.  In the supplementary information, we calculate the critical exponents of the BP model, where in particular it is stressed the difference between the behavior of the mean cluster size of the corona clusters, which diverges with an exponent $\gamma' =1/2$, and the fluctuation of the percolation order parameter, which diverges with an exponent $\gamma=1$.

\section*{Results}
\subsection*{Kinetic facilitated models and bootstrap percolation}

\begin{figure}
\centering
\includegraphics[width=6cm]{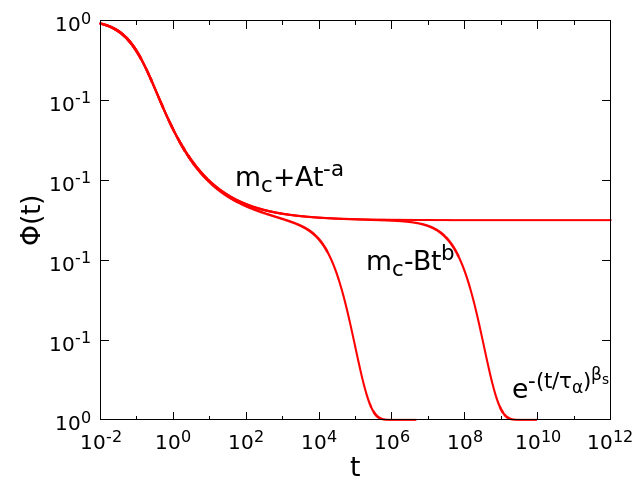}
\caption{Correlators in MCT schematic model B \cite{nota_mct}.
Each curve corresponds to a different value of $\epsilon$. The correlator reaches the plateau with a power law decay, $\Phi(t)-m_c \sim t^{-a}$, and the 
departure from the plateau is given by $\Phi(t) -m_c \sim -\left(t/\tau_\beta\right)^b$. At long times, a crossover is observed to a new regime, well fitted 
by stretched exponential function.}
\label{fig1}
\end{figure}

Kinetic facilitated models \cite{sollich} like Fredrickson and Andersen \cite{FA} or Kob and Andersen models~\cite{KA} on the Bethe lattice 
have been suggested~\cite{sellitto} to have a discontinuous  MCT-like 
transition (see Fig.~\ref{fig1})~\cite{gotze1,gotze2}.
%The correlator $\Phi(t)=\langle q(t)\rangle$ is a measure of how much the system configuration at time $t$
%is correlated to the configuration at time $t=0$.
%For a lattice system, $\Phi(t)$ is often defined as the persistence, with
%\begin{equation}
%q(t)=\frac{1}{N}\sum_{i} n_i(t),
%\label{eq2}
%\end{equation}
%where $n_i(t)=0,1$ depending whether a particle at site $i$ has moved or not during time interval $(0,t)$,
%and $N$ is the total number of particles.
%Typical  behaviour of $\Phi(t)$ is given in  Fig.\ref{fig1}:
%\begin{equation}
%\lim_{t\to\infty}\Phi(t)=m,
%\label{eq25}
%\end{equation}
%at high $T$ (liquid) $m=0$, at low $T$ (glass) $m>0$. 
%At the glass transition the order parameter $m$ jumps discontinuously from zero to a finite value $m_c$, with  $m-m_c \sim \epsilon^\beta$. 
%Close to the transition, in the liquid phase, a two-step relaxation is observed. 
%In the first step, the correlator reaches a plateau with a characteristic time $\tau_\beta\sim \epsilon^{-z_1}$ and, after a departure from the 
%plateau, it approaches zero with a characteristic time $\tau_\alpha\sim\epsilon^{-z}$ 
Our objective is to use a physical picture to understand the origin and the mechanism leading to such peculiar  dynamical behavior. In order to do so, we 
consider, in particular, the Fredrickson and Andersen (FA) kinetic facilitated model~\cite{FA} (FA) on a Bethe lattice. 
The FA model is defined on a lattice, where an Ising variable, $S_i=\pm 1$, is assigned to each of the $N$ sites, with Hamiltonian,
${\cal H}=-\frac{1}{2}\sum_{i=1}^N S_i$.
The spins variables are updated according to the standard spin flip dynamics, along with the constraint that a spin can only flip if the number of nearest
neighbors in the down state ($S_i=-1$) is larger than or equal to $f$. 

The dynamics of the system can be characterized by the correlator, $\Phi(t)=\langle q(t)\rangle$, a quantitative measure of how
the system configuration at time $t$ is correlated to the configuration at time $t=0$, 
the dynamical susceptibility, 
$\chi_4(t)=N (\langle q^2(t)\rangle -­\langle q(t)\rangle^2)$, and the dynamical pair correlation function, $g_{ij}(t)$, where
%The correlator $\Phi(t)$ is a quantitative measure of how 
%the system configuration at time $t$ is correlated to the configuration at time $t=0$.
%For the FA model, $\Phi(t)$ is often defined as the persistence, with
\begin{eqnarray}
q(t)=\frac{1}{N}\sum_{i} n_i(t), ~~~~~~~ g_{ij}(t)=\langle n_i(t) n_j(t)\rangle -\langle n_i(t)\rangle\langle n_j(t)\rangle, ~~~~~~~ \chi_4(t)=
\frac{1}{N}\sum_{ij} g_{ij}(t),
\label{eq2}
\end{eqnarray}
with $n_i(t)=0,1$ depending 
whether a spin at site $i$ has flipped or not during time interval $(0,t)$,
respectively \cite{pastore}.

%and $N$ is the total number of particles.

On a Bethe lattice~\cite{nota1} of coordination number  $z=k+1$, the model,  for $0<f<k-1$, has a transition from a liquid phase at high 
temperatures (where the density of down spins is large), to a frozen phase 
at low temperatures, where down spins are few and an infinite cluster of blocked
spins appears~\cite{toninelli,mendes2006,mendes2008}. It was shown~\cite{sellitto} that 
in the $t\to\infty$ limit, this transition corresponds exactly to that of BP.  
%namely $m = P$ and $ \chi_4(\infty) = \chi$, where $P$ is the bootstrap percolation order 
%parameter and $\chi$ is the fluctuation of $P$ with respect to the initial conditions. 
Bootstrap percolation has a mixed order transition: while the percolation order 
parameter $P$ of BP jumps discontinuously at the threshold from zero to $P_c$, the fluctuation $\chi$ of the order parameter with respect to the initial configuration,  and the associated length $\xi$ diverge according to:
\begin{equation}
P-P_c\sim\epsilon^{\beta},\qquad \chi\sim\epsilon^{-\gamma},\qquad \xi\sim\epsilon^{-\nu},
\label{eq34}
\end{equation}
where 
%\cite{CLR,schwarz,baxter2011,nota2}
\begin{equation}
\beta=1/2,\qquad \gamma=1,\qquad \nu=1/4.
\label{eq35}
\end{equation}

The behavior of the order parameter characterized by the exponent $\beta =1/2$ was first derived in the
original paper where BP was first introduced \cite{CLR}, 
Note that the  fluctuation of the order parameter $\chi$  must not be confused  with the mean cluster size of the ``corona'' . These are clusters made of sites belonging to the percolating cluster, surrounded by a number of facilitated sites exactly equal to $f-1$\cite{schwarz}.
The mean cluster size of the corona in fact diverges with an exponent $1/2$ \cite{schwarz,baxter2011}.
In Ref.\cite{schwarz} a second correlation function was introduced leading to a second ``susceptibility" diverging with an exponent $1$, but it is not clear whether this quantity is related to the fluctuation of the percolation order parameter. In the Supplementary Information, we calculate explicitly the fluctuation of the order parameter and show that it diverges with an exponent $\gamma=1$ along with the associated correlation length $\nu=1/4$.

In summary, for the FA  model we have, in the glassy phase, $m = P$, $\chi_4(\infty) = \chi$,  and $\xi_4(\infty)=\xi$, where $m = \lim_{t\to\infty}\Phi(t)$, 
$\chi_4(\infty) = \lim_{t\to\infty}N(\langle q(t)^2\rangle-\langle q(t)\rangle^2)$ are the FA order parameter and its fluctuation, respectively, and   
$\xi_4(\infty)$ is the associated length. 
%parameter and $\chi$ is the fluctuation of $P$ with respect to the initial conditions. 

{\it Decay to the plateau using the cluster approach.} 
We note that coming from the glassy phase $T<T_c$,  the static properties of the FA model exhibits a mixed order transition at the critical temperature $T_c$, whose critical behavior is given by Eq.~(\ref{eq34}). However,
by re-defining the order parameter as $m-m_c$, the transition can be considered as a continuous one.  Therefore, we can apply the cluster formalism developed for the continuous transition, such as the sol-gel transition and the dynamical transition of the MCT 
model A~\cite{fierro,arenzon}.

%In the sol-gel transition, which is a typical example of continuous transition, 
%using a cluster approach, it was derived \cite{fierro} an explicit scaling form for the correlator $\Phi(t)$ and scaling laws,
%which connect dynamical exponents with the random percolation exponents.
%Recently, it was also shown \cite{arenzon} that the same scaling form with mean field random percolation exponents was also
%valid for the continuous transition of MCT model A, clarifying the origin of the MCT scaling relation and
%suggesting that the same cluster mechanism was responsible for the dynamical transition in both gels and MCT model A.
In the cluster approach, it is assumed that the system can be described by a distribution
of clusters $n(s)$, where each cluster of size $s$ decays with a simple exponential
\begin{equation}
\phi_s(t)\sim e^{-t/\tau_s},
\label{eq8}
\end{equation}
where $\tau_s$ is the relaxation time of a cluster of size $s$. The larger the size of the cluster,
the larger is the relaxation time. It is natural to assume the following power law relation,
as usually found for polymer systems:
\begin{equation}
\tau_s\sim s^x
\label{eq9}
\end{equation}
where $x$ is a constant exponent. The density correlator of the entire system is given by the sum over all clusters
\begin{equation}
\Phi(t) -m \sim \sum_s sn(s)e^{-t/\tau_s}
\label{eq10}
\end{equation}
 where
\begin{equation}
n(s)\sim s^{-\tau}e^{-s/s^\ast}
\label{eq11}
\end{equation}
is the cluster distribution associated to the fluctuation of BP with 
$\tau = 2+\beta/(\beta +\gamma)$ and $s^\ast = \epsilon^{-(\beta + \gamma)}$, where $\beta=1/2$ and $\gamma = 1$ are the mean field BP 
exponents. In the sol-gel transition and MCT model A, the cluster distribution is  given by random percolation theory with $\beta =1$ and $\gamma=1$~\cite{stauffer}. 
Note that this approach is rather general, it is based only under the assumption that the system
configuration can be partitioned in a distribution of clusters, each decaying with a relaxation time proportional
to $s^{x}$.  Even if we do not know precisely the cluster definition, the approach is still valid, just like in a liquid-gas transition 
close to the critical point it is appropriate to describe the critical properties in terms of a
distribution of droplets, in the spirit of Fisher's droplet model~\cite{fisher}.

Provided that we are in the glassy phase, $T\le T_c$, we can apply the cluster formalism of the continuous transition, 
which predicts a pure power law decay~\cite{fierro,arenzon} for the 
entire range of times at the transition, $T=T_c$, and the same power law below  $T_c$, provided that $t\ll\tau_\beta$,
%We note that below the critical temperature the static properties of the FA model, whose critical behaviour is given by Eq.(\ref{eq34}),  are the same as a continuous transition, provided that the order parameter is defined as $P-P_c$. 
%We can therefore associate a cluster distribution to the critical fluctuations,  even if we do not have at the moment a precise definition of 
%clusters.
%At this point, we can apply the cluster formalism of the continuous transition, which predicts a pure power law decay \cite{fierro,arenzon} for the 
%entire range of time at the transition, and the same power law both below and above  $T_c$, provided that $t\ll\tau_\beta$
% 

\begin{equation}
\Phi(t)-m\sim t^{-a},\qquad t\ll\tau_\beta,\qquad\tau_\beta\sim \epsilon^{-z_1}
\label{eq36}
\end{equation}
\begin{equation}
a=\frac{1}{x}\frac{\beta}{\beta+\gamma},\qquad z_1=x(\beta+\gamma)=\frac{\beta}{a},
\label{eq37}
\end{equation}
where $x$ is related to the relaxation time of a fluctuation of size $s$ by Eq.~(\ref{eq9}).
%\begin{equation}
%\tau_s\sim s^x.
%\label{eq38}
%\end{equation}
Inserting BP mean field exponents $\beta=1/2$ and $\gamma=1$, we obtain
\begin{equation}
a=\frac{1}{3x},\qquad z_1=\frac{3x}{2}=\frac{1}{2a}.
\label{eq39}
\end{equation}

Moreover, as in the continuous case,  close to $T_c$ the power law is followed by a transient, whose
behavior is given by a stretched exponential combined with a power law~\cite{fierro,arenzon}: 
\begin{equation}
\Phi(t)-m\sim \epsilon^{\beta}\left(\frac{\tau_\beta}{t}\right)^c e^{-(t/\tau_\beta)^y},
\label{eq42}
\end{equation}
with
\begin{equation}
c=\frac{3\beta+\gamma}{2(x+1)(\beta+\gamma)}=\frac{5}{6(x+1)}=\frac{5a}{2(1+3a)}
\label{eq43}
\end{equation}
and 
\begin{equation}
y=\frac{1}{x+1}=\frac{3a}{1+3a}
\label{eq44}
\end{equation}
\begin{equation}
\frac{c}{y}=\frac{3\beta+\gamma}{2(\beta+\gamma)}=\frac{5}{6}
\label{eq45}
\end{equation}

%Note that the behaviour given by Eq.(\ref{eq42}) can be more evident in the glassy phase, below $T_c$, and it is instead much less pronounced in the liquid phase, 
%above $T_c$, where the regime corresponding to the departure from the plateau will become dominant and interfere with it.

\begin{figure}
\centering
\includegraphics[width=6cm]{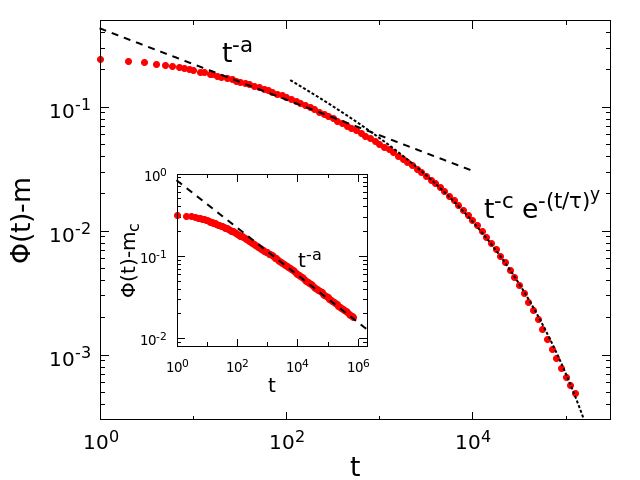}
%{fig3-bis.png}
\caption{The correlator of the FA model on the Bethe lattice in the glassy phase at $T_c-T=2^{-7}$ (main frame) and at the transition (inset). 
The dashed line is the power law $t^{-a}$ with exponent $a= 0.29$.
The dotted line in the glassy phase is given by Eq.~(\ref{eq42}), with
$y= 0.46$ and  $c=0.39$. 
}
\label{fig2}
\end{figure}

We have performed large scale numerical simulations of the FA model on the Bethe lattice with $k=3$, $f=2$ and $N=2^{18}$. 
Fig.~\ref{fig2} shows the correlator in the glassy phase (main frame) and at the transition (inset). 
The value of $a\approx 0.29$ is obtained from the power law decay at the critical 
temperature. From this value, using Eqs.~(\ref{eq43}) and (\ref{eq44}), we predict 
the exponents $y\approx 0.46$ and $c\approx 0.39$, defined in Eq.~(\ref{eq42}).
Fig.~\ref{fig2} shows that the data  are in excellent agreement
with the cluster approach predictions.

In the liquid phase,  we have the same approach to the plateau, Eq.~(\ref{eq36})
replacing $m$ by $m_c$, and the same power law, Eq.~(\ref{eq37}), provided that 
the system is close enough to $T_c$ and $t\ll\tau_\beta$. 
Note that the behavior given by Eq.~(\ref{eq42})  is much less pronounced in the liquid phase, since for large $t$ the regime corresponding to the departure from the plateau will become dominant and interfere with it.  

{\it Departure from the plateau using the damage spreading mechanism.}
In the liquid phase, all the clusters (fluctuations) vanish in the long time limit, but they survive on time scales 
of the order of $\tau_{\beta}$, when the plateau is still present. The small clusters start to decay first, the last clusters to relax are the largest clusters, i.e. the critical clusters. Once the sites in the critical clusters have moved (relaxed), they act as initial damaged sites to ``free'' the sites of the potential bootstrap percolating cluster represented by the plateau. As the time increases, the damage  spreads through a  branching cascade process~\cite{mendes2015}. The physical picture behind it is that the potential infinite cluster, which contributes to the plateau, is made of a sea of quasi frozen sites, surrounded by critical clusters. Just above the critical temperature the critical clusters eventually decay, whereas just below the critical temperature, the critical clusters themselves become frozen and part of the infinite cluster.  
The number of sites $m(t)$ in the core, which are liberated by the damage spreading, is related to the correlator by
\begin{equation}
\Phi(t)\simeq m_c-m(t),
\label{eq47}
\end{equation}
\begin{equation}
m(t)\sim\epsilon^\beta\left(t/\tau_\beta\right)^b,\qquad\beta=1/2,
\label{eq48}
\end{equation}
where $\epsilon^\beta$ is the density of sites in the critical clusters and therefore the density of initial damaged sites, and 
$1/\tau_\beta$ is, according to the cluster picture, the diffusion coefficient of the sites in the critical cluster, and
$b$ is a dynamical exponent related to the spreading damage mechanism.
%In presence of correlation a value of $b\ne 1$ is expected \cite{toninelli2,pastore}, being instead $b=1$ in 
%the random walk. 

\begin{figure}
\centering
\includegraphics[width=6cm]{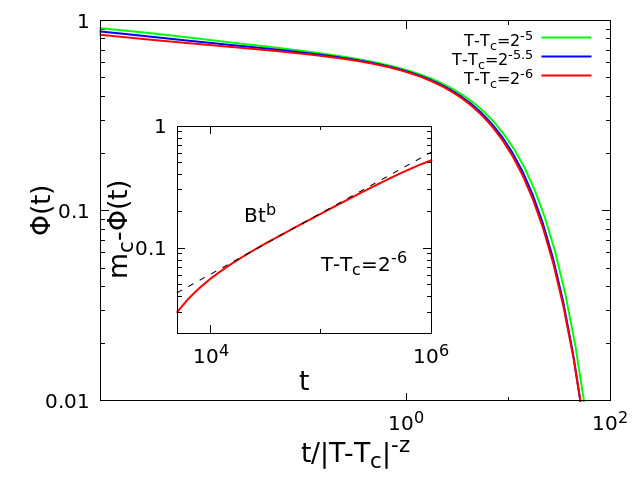}
\caption{
Main frame: Scaling collapse of the 
correlator of the FA model on the Bethe lattice, in
the liquid phase $T>T_c$, Eq. (\ref{eq41}), with $z=2.72$.
Inset: Departure from the plateau,
the dashed line is the power law with $b=0.50$.
%, that satisfy the MCT
%relation (21) with $\lambda=0.785$.
%Data collapse for the correlator of the FA model on the Bethe lattice with $k=3$ and $f=2$, in the liquid phase $T>T_c$.
%The critical density is $p_c=8/9$, corresponding to $T_c= 0.480$,
%and the fraction of blocked spins at the critical point is $m_c= 0.673$.
%The fitting exponents are $a\simeq 0.29$ and $b\simeq 0.50$, that satisfy the MCT 
%relation Eq.~(\ref{eq52a}) with $\lambda\simeq 0.785$.
}
\label{fig3}
\end{figure}

Finally, in the $\alpha$ regime, 
\begin{equation}
\Phi(t)\simeq g\left(t/\tau_\alpha\right),\qquad \tau_\alpha\sim\epsilon^{-z},
\label{eq41}
\end{equation}
and, like in MCT, using the matching conditions with the previous regime
\begin{equation}
z=\frac{\beta}{a}+\frac{\beta}{b}=\frac{1}{2a}+\frac{1}{2b}.
\label{eq51}
\end{equation}
In Fig.~\ref{fig3}, 
we have reported the scaling collapse of the correlator in the $\alpha$ regime, Eq.~(\ref{eq41}), from which   
the exponent $z\approx 2.7$ of the relaxation time $\tau_\alpha$ has been evaluated.This value  is consistent with the value found in Ref.~\cite{sellitto}. In the inset we have also reported the departure from the plateau Eq.~(\ref{eq48}) and the value of $b\approx 0.50$ has been evaluated.
The exponents $z$, $a$ and $b$ satisfy not only the scaling relation ~(\ref{eq48}), but  $a$ and $b$ are also found to satisfy the other MCT relation, Eq. (\ref{eq52a}), with $\lambda=0.785$.

%\textcolor{magenta}{Unfortunately, at the moment, it is unclear if such a crossover can be predicted quantitatively as in MCT \cite{maurocrossover}, by the damage spreading mechanism. To make further progress in this direction one should assume (quite reasonably?) that corrections to Eq.(\ref{eq48}) do have the form of a Taylor series in the variable $(t/\tau_{\beta})^b$ and similarly for the approach to the plateau (with $b\rightarrow -a$).}
%\textcolor{magenta}{
%we have reported the data collapse for the correlator around the plateau in the time window of 
%the crossover between the  $\beta$ and $\alpha$ relaxation, Eq.(\ref{eq29}), where the value of $b\approx 0.48$ have been evaluated \cite{nota4}.
%}

Given two critical times $\tau_\beta$ and $\tau_\alpha$,in the liquid phase it may be more convenient to express the density correlator $\Phi(t)$ as a scaling function 
of two variables: 
\begin{equation}
\Phi(t) -m_c \simeq \epsilon^{\beta}F\left(\frac{t}{\tau_\beta},\frac{t}{\tau_\alpha}\right),
\label{eq29}
\end{equation}
$F(x,y) = F_1(x) $ for $y \ll1$, 
where %$F_1(x)$ behaves as
$F_1(x) = x^{-\beta/z_1}$  for $x \ll1$  and $F_1(x) = - x^{b}$  for $x>1$,
and  $F(x,y)  = - x^b F_2(y)$   for $x\gg 1$ and $y>1$.  
%\textcolor{magenta}{$F_1(x) = - x^{b}$  for $x\gg 1$ and $y\ll 1$,
%and  $F(x,y)  = - x^b F_2(y)$   for $x\gg 1$ and $y\gg 1$.}
The requirement that $\Phi(t)$ for $t>\tau_\alpha$ %\textcolor{magenta}{$t\gg \tau_\alpha$}  
is a function of $t/\tau_\alpha$ only, 
Eq.~(\ref{eq41}), implies that 
$\epsilon^{\beta}x^b=y^b$, which in turn leads to $\tau_\alpha= \tau_b \tau_{\beta}$ where 
$\tau_b\sim \epsilon^{-z_b}$ with $z_b= 1/2b$. Taking into account that $\tau_{\beta} \sim \epsilon^{-z_1}$, 
it follows the scaling relation Eq.~(\ref{eq51}).
\subsection*{Comparison with discontinuous MCT model B}

Interestingly, the correlator of the MCT model B \cite{nota_mct} satisfies the same scaling forms 
Eqs.~(\ref{eq36}), (\ref{eq48}), (\ref{eq41}) and 
scaling relations Eqs.~(\ref{eq37}) and (\ref{eq51}), suggesting that 
the above picture is consistent with MCT.
This is further validated if we consider that the mean field static BP exponents coincide with those found in the 
Random Field Ising (RFI) model in an external 
field~\cite{RFI1,RFI2,RFI3,birolitarjus}, which was
shown to be mapped on the MCT theory, and that both BP and RFI model have an upper critical dimension 
$d_c = 8$, which coincides with the value found for MCT~\cite{BBdc}.  

MCT also predicts a relation between the exponent $a$ and $b$ and the MCT parameter $\lambda$:
\begin{equation}
\frac{\Gamma^2(1-a)}{\Gamma(1-2a)} =\frac{\Gamma^2(1+b)}{\Gamma(1+2b)} = \lambda.
\label{eq52a}
\end{equation}
In our approach instead of $\lambda$ we have $x$ as parameter, which is related to $a$ 
through
\begin{equation}
a=\frac{1}{x}\frac{\beta}{\beta+\gamma}=\frac{1}{3x}.
\label{eq52b}
\end{equation}
If our  approach applies to MCT, $x$  can be related to $\lambda$ and  consequently to $b$ through Eq.~(\ref{eq52a}) 
\begin{equation}
\frac{\Gamma^2(1+b)}{\Gamma(1+2b)}=\frac{\Gamma^2(1-1/3x)}{\Gamma(1-2/3x)}=\lambda.
\label{eq52c}
\end{equation}

At the moment, we do not have an intuitive physical picture of
why $x$ and $b$ should be related in such a manner. Interestingly,  our data show that the MCT relation  
Eq.~(\ref{eq52a}) is  well verified on the FA model, strongly supporting the idea that the FA model in mean field reproduces entirely MCT~\cite{ArSe,FrSe}.   

\begin{figure}
\centering
\includegraphics[width=6cm]{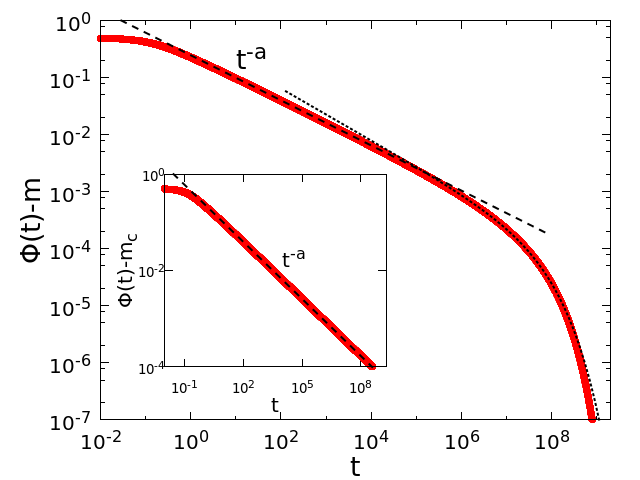}
\caption{Numerical solution of the MCT schematic model B in the glassy phase (red continuous line), with $\lambda=0.5$.
Dashed line is the power law $t^{-a}$, with $a= 0.40$, obtained from Eq.~(\ref{eq52a}).
The data are in excellent agreement with the cluster approach prediction
(dotted line is the stretched exponential-like decay, Eq.(\ref{eq42}), with $c=0.45$ and $y=0.54$,
obtained using $x=1/3a= 0.84$).}
\label{fig4}
\end{figure}

Our approach predicts that the approach to the plateau, after a power law behavior, should be described by the stretched exponential Eq.~(\ref{eq42}) with exponents given 
by Eqs.~(\ref{eq43}) and (\ref{eq44}), before becoming an exponential decay.
By numerically solving the MCT schematic model, we found that the approach to the plateau is very well described by the above predictions (see Fig.~\ref{fig4}).

\subsection*{Fluctuations of the order parameter}

Dynamical heterogeneities play an important role in understanding the nature of the glass transition
\cite{het1,het2,het3,het4,het5,het6,het7,het8,het9,het10,het11,het12}.
They are described through the dynamical susceptibility, $\chi_4(t)$,
defined as the fluctuations of the dynamical order parameter:
$\chi_4(t)=N(\langle q(t)^2\rangle-\langle q(t)\rangle^2)$.
%\label{eq54}
In the following, we will refer to the FA model, however the same predictions apply to the MCT model as well, if the two models behave in the same way, as shown already for the decay of the correlator.
As for the correlator, we express $\chi_4(t)$ as a scaling function of two variables. Since in the glassy phase 
for t going to infinity $\chi_4(t)$ coincides with the fluctuation of the BP order parameter, which diverges with an exponent $\gamma=1$ as the glass transition is approached we can write:
\begin{equation}
\chi_4(t)\simeq \epsilon^{-\gamma}G_{\pm}\left(\frac{t}{\tau_\beta},\frac{t}{\tau_\alpha}\right).
\label{eq55}
\end{equation}
where $G_{\pm}(x,y)$ is a two variables scaling function in the glassy ($+$) and liquid ($-$) phase.

{\it Glassy phase.}
Given that in the glassy phase $\tau_\alpha=\infty$,  we have
\begin{equation}
\chi_4(t)\simeq\epsilon^{-\gamma}F_{+}\left(\frac{t}{\tau_\beta}\right).
\label{eq56}
\end{equation}
where $F_{+}\left(\frac{t}{\tau_\beta}\right)=G_{+}\left(\frac{t}{\tau_\beta},0\right)$. $F_{+}=\text{const}$ for   $t\rightarrow \infty$ and $F_{+}\left(\frac{t}{\tau_\beta}\right)=\left(\frac{t}{\tau_\beta}\right)^{a\gamma/\beta}$ for~~ $t<\tau_\beta$,
%$t=\tau_\beta$,  $\chi_4\sim\epsilon^{-\gamma} =\epsilon^{-1}$, 
so that since $\tau_\beta\sim\epsilon^{-\beta/a}$, 

\begin{equation}
\chi_4(t)\sim t^{a\gamma/\beta} = t^{2a}\qquad\mbox{for~~ $t<\tau_\beta$,}
\label{eq57}
\end{equation}
 is independent on $\epsilon$, and 
\begin{equation}
\chi_4(t)\sim  \epsilon^{-\gamma} =\epsilon^{-1} \qquad\mbox{for~~ $t\rightarrow \infty$,}
\label{eq57b}
\end{equation}
In conclusion in the glassy phase $\chi_4(t)$ grows as a power law $t^{2a}$ until it reaches a plateau at $t\sim\tau_\beta$,  whose value diverges with an exponent $\gamma=1$.
 
{\it Liquid phase.}
In the liquid phase,  we have:
%\begin{equation}
%\chi_4(t)\simeq\epsilon^{-\gamma}G_{-}\left(\frac{t}{\tau_\beta},\frac{t}{\tau_\alpha}\right),
%\label{eq58}
%\end{equation}
\begin{equation}
\chi_4(t)\simeq\epsilon^{-\gamma}F_{-}\left(\frac{t}{\tau_\beta}\right)\qquad\mbox{for $t\ll\tau_\alpha$.}
\label{eq59}
\end{equation}
where $F_{-}\left(\frac{t}{\tau_\beta}\right)=G_{-}\left(\frac{t}{\tau_\beta},0\right)$.
In the early regime, $t\le\tau_\beta$, the behavior is the same as in the glassy phase,
\begin{equation}
\chi_4(t)\sim t^{a\gamma/\beta} = t^{2a}\qquad\mbox{for~~ $t<\tau_\beta$,}
%\chi_4(t) F_{-}\left(\frac{t}{\tau_\beta}\right)=\left(\frac{t}{\tau_\beta}\right)^{a\gamma/\beta}\qquad\mbox{for $t<\tau_\beta$,}
\label{eq61}
\end{equation}

\begin{equation}
\chi_4\sim\epsilon^{-\gamma}=\epsilon^{-1}\qquad\mbox{for $t=\tau_\beta$,}
\label{eq60}
\end{equation}

%\begin{equation}
%F_{-}=\text{const.},\qquad\chi_4\sim\epsilon^{-\gamma}=\epsilon^{-1}\qquad\mbox{for $t=\tau_\beta$,}
%\label{eq60}
%\end{equation}

%so that, since $\tau_\beta\sim\epsilon^{-\beta/a}$, $\chi_4(t)\sim t^{a\gamma/\beta} = t^{2a}$ is independent on $\epsilon$.

\begin{figure}
\centering
\includegraphics[width=6cm]{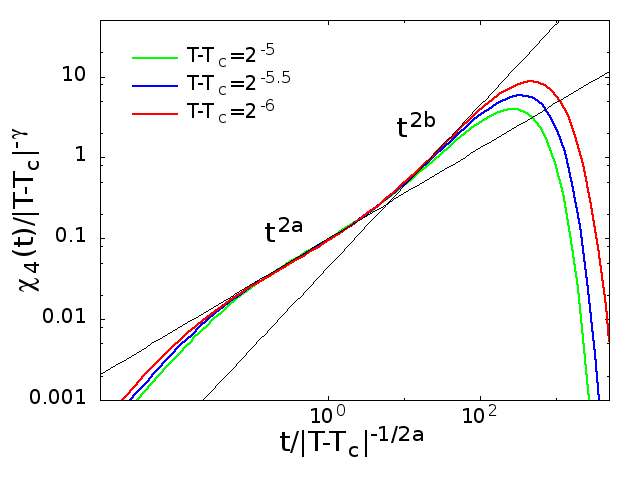}
\caption{Data collapse in the $\beta$ regime of the dynamical susceptibility, $\chi_4(t)$,  for the FA model on the Bethe lattice, 
showing the scaling relation Eq.~(\ref{eq59})
with $\gamma=1$, $\tau_\beta\sim \epsilon^{-1/2a}$.
Straight lines show the power law behaviors in the
early ($t^{2a}$) and late $\beta$
regime ($t^{2b}$), with $a= 0.29$ and $b=0.50$.}
\label{fig5}
\end{figure}

We have numerically verified on FA model on the Bethe lattice
the scaling relation Eq.~(\ref{eq59}). In Fig.~\ref{fig5}, we have reported the rescaled susceptibility, showing that 
in the $\beta$ regime $t\ll \tau_\alpha$ all curves rescale onto a unique function corresponding to $F_-(x)$ (with $x=t/\tau_\beta$), 
and that $\chi_4(t) \sim \epsilon^{-\gamma}$  for $t=\tau_\beta$ (with  $\gamma = 1$).

In the late $\beta$ regime, $\tau_\beta<t\ll\tau_\alpha$, as the dynamical process is due to the damage spreading mechanism,  
$\chi_4(t)$ must be  proportional to the square of the number of visited sites $m^2(t)\sim t^{2b}$ \cite{conigliodamage} 
similar to what is  found  in the diffusing defects mechanism (see 
Refs.~\cite{toninelli2,pastore}), therefore 
\begin{equation}
F_{-}\left(\frac{t}{\tau_\beta}\right)=\left(\frac{t}{\tau_\beta}\right)^{2b}\qquad\mbox{for $\tau_\beta<t\ll\tau_\alpha$}
\label{eq62}
\end{equation}

From Eqs.~(\ref{eq59}) and (\ref{eq62})
\begin{equation}
\chi_4(t)\sim\epsilon^{-\gamma-2\beta}\left(\frac{t}{\tau_\alpha}\right)^{2b}\qquad\mbox{for $\tau_\beta<t\ll\tau_\alpha$},
\label{eq63}
\end{equation}
where $\tau_\alpha\sim\epsilon^{-z}$,  $\tau_\beta\sim \epsilon^{-\beta/a}$ and the scaling relation Eq.~(\ref{eq51}) has been taken into account.

\begin{figure}
\centering
\includegraphics[width=6cm]{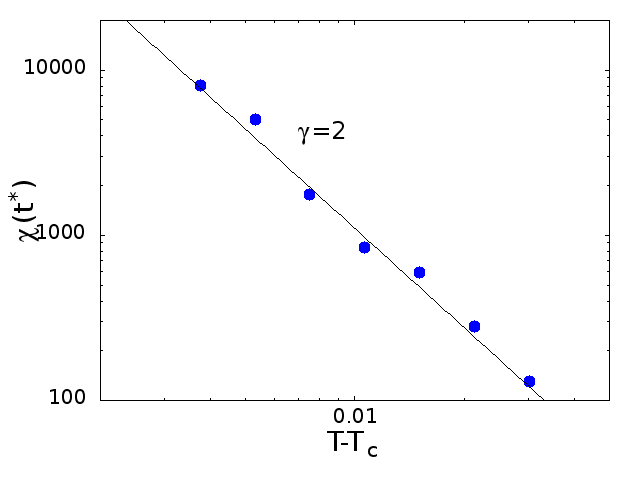}
\caption{Maximum of the dynamical susceptibility $\chi_4(t^*)$  for the FA model on the Bethe lattice as a function of $T-T_c$. The maximum diverges as $|T-T_c|^2$, in agreement with
the prediction of Eq.~(\ref{eq65}).}
\label{fig6}
\end{figure}

In general, in the late $\beta$ and $\alpha$ regime $t>\tau_\beta$, from Eq.~(\ref{eq59}) we have 
$G_{-}\left(\frac{t}{\tau_\beta},\frac{t}{\tau_\alpha}\right) =\left(\frac{t}{\tau_\beta}\right)^{2b} H_{-}\left(\frac{t}{\tau_\alpha}\right)$, where $H_{-}(y) = \text{const.}$, for $y\le 1$ in order to match the behavior in the late $\beta$ regime Eq.~(\ref{eq62}), and goes to zero for $y>>1$, as $\chi_4(t)$ in the infinite time limit tends to the value of the BP susceptibility $\chi$, which is zero in the liquid phase. Therefore
\begin{equation}
\chi_4(t)=\epsilon^{-\gamma-2\beta}\left(\frac{t}{\tau_\alpha}\right)^{2b}H_{-}\left(\frac{t}{\tau_\alpha}\right)\qquad\mbox{for $\tau_\beta<t$},
\label{eq64}
\end{equation}
%where $H_{-}(y)=\text{const.}$ for $y=1$, 
\begin{equation}
\chi_4\sim\epsilon^{-\gamma-2\beta}\sim\epsilon^{-2}\qquad\mbox{for $t=\tau_\alpha$,}
\label{eq65}
\end{equation}
where $\beta=1/2$, $\gamma=1$, have been taken into account. We have found good agreement for  the FA model as shown in Fig.~\ref{fig6}, where the maximum of $\chi_4(t)$ is plotted for $t=t^*\sim\tau_\alpha$, and in  Fig.~\ref{fig5},  where it is shown  $\chi_4(t) \sim t^{2a }$ in the early $\beta$
regime and $\chi_4(t)\sim t^{2b}$ in the late $\beta$ regime.

%\begin{figure}
%\centering
%\includegraphics[width=6cm]{fig7.png}
%\caption{Dynamical susceptibility of the FA model on the Bethe lattice  for different temperatures.
%Straight lines show the power law behaviors in the
%early ($t^{2a}$) and late $\beta$  
%regime ($t^{2b}$).}
%\label{fig7}
%\end{figure}

{\it Comparison with MCT-}  $\chi_4(t)$ was studied within the $p$-spin model by Franz and Parisi~\cite{het4} and within the MCT theory by Biroli and Bouchaud~\cite{BB} using a diagrammatic approach. The MCT results~\cite{toninelli2} predicted for $\chi_4(t)$ a growth  respectively $t^a$ and $t^b$ for the early and late $\beta$ regime and a growth of the maximum at $t^*$ with an exponent $1$. Later it was argued~\cite{berthier}  that this behavior is valid only for ensembles where all conserved degrees of freedom are fixed, e.g. Newtonian dynamics in the NVE ensemble or Brownian dynamics, in the NVT ensemble, otherwise other diagrams would contribute to $\chi_4(t)$ leading to a behavior  $t^{2a}$   and $t^{2b}$ and an exponent $2$ for the growth of the maximum of $\chi_4(t)$ at $t^*$. The same found in our approach. More recently in~\cite{franz}, this last behavior was found to be much more general,  being due to self induced disorder. Changing the initial condition induces fluctuation in the induced 
disorder leading to the new dynamical behavior. 

\section*{Discussion} 

%In conclusion, we have shown that a cluster approach and a damage spreading mechanism, applied to the FA kinetic facilitated model
%in mean field, predict discontinuous dynamical transition with the same scaling behavior found in the discontinuous MCT transition 
%and Random Field Ising model in a field. 
%The dynamical transition is characterized by a static mixed order transition in the same universality class 
%of  BP. 
%This scenario at least at the mean field level suggests that the dynamical transition occurs together with a static critical point, which is manifested coming from the glassy phase. Although the static transition is absent in the liquid phase, the dynamics in the liquid phase is influenced by the static transition in the glassy 
%phase. 
%The presence of the static transition, at least at mean field level, characterized by a diverging static length, is responsible for 
%the scaling laws and universality present in a wide range of dynamical critical phenomena, and sets the value of the upper 
%critical dimensionality $d_c=8$. In this scenario the sol-gel transition, which in mean field has been shown to be described by 
%the continuous MCT model A \cite{arenzon}, can be considered as  dynamical transition in a different universality class, 
%characterized by the static random percolation transition  with upper critical dimensionality $d_c=6$. 

In conclusion, we have shown that a cluster approach and a damage spreading mechanism, applied to the FA kinetic facilitated model
in mean field, predict a discontinuous dynamical transition with the same scaling behavior found in the discontinuous MCT transition 
and Random Field Ising model in a field. 
The dynamical transition is characterized by a static mixed order transition, in the same universality class 
of bootstrap percolation. This static transition is characterized by static critical fluctuations  diverging only in the glassy phase, being absent in the liquid phase. Nevertheless the dynamics even in the liquid phase is strongly influenced by this static transition, as  shown by the behavior  of the dynamical heterogeneities, characterized by $\chi_4(t)$. 
%This scenario at least at the mean field level suggests that the dynamical transition occurs together with a static critical point, which is manifested coming from the glassy phase. Although the static transition is absent in the liquid phase, the dynamics also in the liquid phase is influenced by the static transition in the glassy 
%phase. 
The presence of the static transition, at least at mean field level, characterized by a diverging static length, is responsible for 
the scaling laws and universality present in a wide range of dynamical critical phenomena, and sets the value of the upper 
critical dimensionality $d_c=8$. In this scenario the sol-gel transition, which in mean field has been shown to be described by 
the continuous MCT model A \cite{arenzon}, can be considered as  dynamical transition in a different universality class, 
characterized by the static random percolation transition  with upper critical dimensionality $d_c=6$. ,

\section*{Methods}

We performed Monte Carlo simulations of the FA kinetic facilitated model on a random lattice with $N=2^{18}$ sites, fixed coordination number $z=k+1=4$, and $f=2$.
For each temperature we extract 32 different random lattices and initial configurations,
and we start from a random configuration of the spins with density $p=(1+e^{-1/T})^{-1}$ of the up spins.
Each Monte Carlo step is given by $N$ spin flip trials. A spin flip trial consists in taking a random spin, and flipping it if
it has $f$ or more neighboring down spins, and with probabilities given by $p(-\to +)=\frac{e^{1/2T}}{e^{1/2T}+e^{-1/2T}}$ and $p(+\to -)=\frac{e^{-1/2T}}{e^{1/2T}+e^{-1/2T}}$.
For $k=3$ and $f=2$ the critical temperature is $T_c=\left[\ln\left(\frac{p_c}{1-p_c}\right)\right]^{-1}\approx 0.480898$.

The relaxation function is defined as
\[
\Phi(t)=\langle q(t)\rangle
\]
where $q(t)=\frac{1}{N}\sum_{i} n_i(t)$, and $n_i(t)=0,1$ depending
whether a spin at site $i$ has flipped or not during time interval $(0,t)$,
while the fluctuations are defined as
\[
\chi_4(t)=N\left[\langle q(t)^2\rangle-\langle q(t)\rangle^2\right]
\]
where $\langle\cdots\rangle$ is the average over the thermal noise, the initial configurations and the random lattice.

\section*{Acknowledgments} 
The authors would like to thank J. Arenzon,  G. Biroli and M. Sellitto  for many interesting discussions, valuable suggestions and critical reading of the manuscript, and S. Franz, M. Pica Ciamarra, R. Pastore and M. Tarzia for interesting discussions.

We acknowledge financial support from the CNR-NTU joint laboratory {\it Amorphous materials for energy harvesting
applications}.

\section*{Contributions}

A.D. A.F., A.C. conceived the project, A.D. and A.F. carried out
simulations and analyzed the data, A.D. A.F. A.C. wrote the paper.


\begin{thebibliography}{99}
\bibitem{flo}
Flory, P. J. {\it The Physics of Polymer Chemistry}, Cornell
University Press (1954).
\bibitem{deg}
de Gennes, P. G. {\it Scaling Concepts in Polymer Physics}, Cornell
University Press (1993).
\bibitem{zac}
Zaccarelli, E. Colloidal gels: equilibrium
and non-equilibrium routes, {\it J. Phys.: Condens. Matter} {\bf 19}, 323101 (2007)
\bibitem{fierro}
Fierro, A., Abete, T., $\&$ Coniglio, A.
Static and Dynamic Heterogeneities in a Model for Irreversible Gelation,
{\it J. Chem. Phys.} {\bf 131} (19), 194906 (2009).
\bibitem{arenzon}
Arenzon, J. J., Coniglio, A., Fierro, A., $\&$ Sellitto, M.
Percolation approach to glassy dynamics with continuously broken ergodicity,
{\it Phys. Rev. E} {\bf 90}, 020301(R) (2014);
Coniglio, A., Arenzon, J. J., Fierro, A., $\&$ Sellitto, M.
Relaxation dynamics near the sol–gel transition: From cluster approach to mode-coupling theory,
{\it Eur. Phys. J. Special Topics} {\bf 223}, 2297–2306 (2014).
\bibitem{gotze1}
G\"otze, W., $\&$ Sjogren, L. Relaxation processes in supercooled liquids,
{\it Rep. Prog. Phys.} {\bf 55}, 241 (1992).
\bibitem{gotze2}
G\"otze, W. Recent tests of the mode-coupling theory for glassy dynamics, {\it J. Phys. Cond. Matt.} {\bf 11}, A1 (1999).
\bibitem{pusey} Pusey, P. N., $\&$ van Megen, W.
Observation of a glass transition in suspensions of spherical colloidal particles
{\it Phys. Rev. Lett}  {\bf  59}, 2083 (1987).
\bibitem{kob} Kob, W., $\&$ Andersen, H. C.
Testing mode-coupling theory for a supercooled binary Lennard-Jones mixture I: The van Hove correlation function
{\it Phys. Rev. E} {\bf 51}, 4626 (1995).
Kob, W., $\&$ Andersen, H. C.
Testing mode-coupling theory for a supercooled binary Lennard-Jones mixture. II. Intermediate scattering function and dynamic susceptibility
{\it Phys. Rev. E} {\bf 52}, 4134 (1995).
\bibitem{dawson}
Dawson, K., Foffi, G., Fuchs, M., G\"otze, W., Sciortino, F., Sperl, M., Tartaglia, P., Voigtmann, Th., $\&$
Zaccarelli, E.
Higher-order glass-transition singularities in colloidal systems with attractive interactions
{\it Phys. Rev. E}  {\bf 63}, 011401 (2000)
\bibitem{mallamace} Chen, S. H., Chen, W. R., $\&$ Mallamace, F.
The Glass-to-Glass Transition and Its End Point in a Copolymer Micellar System
{\it Science} {\bf 300}, 619 (2003).
Mallamace, F., Corsaro, C., Stanley, H.E., Mallamace, D., Chen, S.H.
The dynamical crossover in attractive colloidal systems
{\it J Chem Phys.}  {\bf 139}, 214502 (2013).
\bibitem{pspin1} M\`ezard, M., Parisi, G., $\&$ Virasoro, M. A. {\it Spin Glass Theory and
Beyond}, World Scientific, Singapore (1987).
\bibitem{pspin2} Cugliandolo, L. F., $\&$ Kurchan, J.
Analytical solution of the off-equilibrium dynamics of a long-range spin-glass model,
{\it Phys. Rev. Lett.} {\bf 71}, 173 (1993).
\bibitem{RFI1} Stevenson, J. D., Walczak, A. M., Hall, R. W., $\&$ Wolynes, P. G.,
Constructing explicit magnetic analogies for the dynamics of glass forming liquids,
{\it J. Chem. Phys.} {\bf 129}, 194505 (2008).
\bibitem{RFI2}
Franz, S., Parisi, G., Ricci-Tersenghi, F., $\&$ Rizzo T.,
Properties of the perturbative expansion around the mode-coupling dynamical transition in 
glasses, arXiv:1001.1746
\bibitem{RFI3} Biroli, G., Cammarota, C., Tarjus, G., $\&$ Tarzia, M., 
Random-field-like criticality in glass-forming liquids,
{\it Phys. Rev. Lett.} {\bf 112}, 175701 (2014).
\bibitem{birolitarjus} Nandi, S., Biroli, G., $\&$ Tarjus, G.,
Spinodals with Disorder: from Avalanches in Random Magnets to Glassy Dynamics, arXiv:1507.06422
\bibitem{FA}
Fredrickson, G. H., $\&$ Andersen, H. C. Kinetic Ising Model of the Glass Transition, {\it Phys. Rev. Lett.} {\bf 53}, 1244 (1984).
\bibitem{KA}
Kob, W. $\&$ Andersen, H. C.
Kinetic lattice-gas model of cage effects in high-density liquids and a test of mode-coupling theory of the ideal-glass transition,
{\it Phys. Rev. E} {\bf 48}, 4364 (1993).
\bibitem{sollich}
Ritort, F. $\&$  Sollich, P.
Glassy dynamics of kinetically constrained models
{\it Advances in Physics} {\bf 52}, 219 (2003).
\bibitem{sellitto}
Sellitto, M., Biroli, G., $\&$ Toninelli, C. 
Facilitated spin models on Bethe lattice: bootstrap percolation, mode coupling transition and glassy dynamics,
{\it Europhys. Lett.} {\bf 69}, 496 (2005).
\bibitem{CLR}
Chalupa, J., Leath, P. L., $\&$ Reich, R. 
Bootstrap percolation on a Bethe lattice, {\it J. Phys. C: Solid State Phys.} {\bf 12}, 
L31 (1979).
%\bibitem{nota3} 
%The bootstrap percolation model \cite{CLR} is defined in the following way. First the sites of the lattice are occupied randomly with probability $p$. Then each site that has $f$ or more empty neighbors is removed, until a stable configuration is reached. The clusters of particles in this configuration  are the bootstrap percolation clusters.
\bibitem{schwarz}
Schwarz, J. M., Liu, A. J., $\&$ Chayes, L. Q. {\it Europhys. Lett.} {\bf 73} 
84, 560 (2006).
\bibitem{baxter2011} 
Baxter, G. J., Dorogovtsev, S. N., Goltsev, A. V. $\&$ Mendes, J. F. F. 
Heterogeneous k-core versus bootstrap percolation on complex networks,
{\it Phys. Rev. E} {\bf 83}, 051134 (2011).
%\bibitem{nota2}
%Note that the clusters must not be confused with the so called ``corona'', which in the glassy phase is part of the frozen infinite cluster,
%made of ``fragile'' sites  surrounded by a number of facilitated sites exactly equal to $f-1$.
%The mean cluster size of the corona in fact diverges with an exponent $1/2$ and not with the $\gamma=1$ related to the fluctuation of the infinite cluster
%(see Ref. \cite{schwarz} and Supplementary information).
\bibitem{ArSe} Arenzon, J.J. $\&$ Sellitto, M. 
Microscopic models of mode-coupling theory: the $F_{12}$ scenario,
{\it J. Chem. Phys.} {\bf 137}, 084501 (2012). 
Sellitto, M., De Martino, D., Caccioli, F. $\&$ Arenzon, J.J. 
Dynamic facilitation picture of a higher-order glass singularity,
{\it Phys. Rev. Lett.} {\bf 105}, 265704  (2010).
\bibitem{FrSe} Franz, S. $\&$  Sellitto, M. 
Finite-size critical fluctuations in microscopic models of mode-coupling theory,
{\it J. Stat. Mech.} P02025 (2013).
\bibitem{maurocrossover}
Sellitto, M. 
Crossover from beta to alpha Relaxation in Cooperative Facilitation Dynamics,
{\it Phys. Rev. Lett.} {\bf 115}, 225701 (2015).
\bibitem{pastore}
Pastore, R., Coniglio, A., Pica Ciamarra, M. Dynamical Correlation Length and Relaxation Processes in a Glass Former
{\it Phys. Rev. Lett.} {\bf 107}, 065703 (2011).
\bibitem{nota1}
The Bethe lattice is a lattice extracted randomly from the set of lattices where each site
is connected to $z=k+1$ other sites. In the following we will consider the FA model on the Bethe lattice with $k=3$ and $f=2$,
that has a transition at a density $p_c=8/9$ of up spins, corresponding to
$T_c=\left[\ln\left(\frac{p_c}{1-p_c}\right)\right]^{-1}\approx 0.480898$, and a
fraction of blocked spins equal to $m_c\approx 0.673$.
\bibitem{toninelli}
Toninelli, C., Biroli, G. $\&$ Fisher, D. S. 
Spatial structures and dynamics of kinetically constrained models for glasses,
{\it Phys. Rev. Lett.} {\bf 92}, 185504 (2004).
\bibitem{mendes2006} 
Dorogovtsev, S. N., Goltsev, A. V. $\&$ Mendes, J. F. F. 
$K-$core organization of complex networks,
{\it Phys. Rev. Lett.} {\bf 96}, 040601 (2006).
\bibitem{mendes2008}  
Dorogovtsev, S. N., Goltsev, A. V., $\&$ Mendes, J. F. F.
Critical phenomena in complex networks,
{\it Rev. Mod. Phys.} {\bf 80}, 1275 (2008).
\bibitem{stauffer}
Stauffer, D., $\&$ Aharony, A. {\it Introduction to Percolation
Theory}, Taylor \& Francis (1992).
\bibitem{fisher} 
%Fisher, M.E. {\it Physics} (N.Y.) {\bf 3}, 225  (1967);
Fisher, M. E. Magnetic Critical Point Exponents$-$Their Interrelations and Meaning,
{\it J. Appl. Phys.} {\bf 38}, 981 (1967);
Fisher, M. E. $\&$ Widom, B.
Decay of Correlations in Linear Systems 
{\it J. Chem. Phys.} {\bf 50}, 3756 (1969);
Fisher, M. E. 
In: {\it Critical Phenomena, Proceedings of the International School of Physics ``Enrico Fermi'' Course LI}, Varenna on lake Como (Italy),
M. S. Green (ed), Academic, New York, p. 1 (1971).
\bibitem{mendes2015} G. J. 
Baxter, G. J., Dorogovtsev, S. N., Lee, K.-E., Mendes, J. F. F.  $\&$ Goltsev, A. V. 
Critical dynamics of the $k-$core pruning process,
{\it Phys. Rev. X} {\bf 5}, 031017 (2015). 
\bibitem{nota_mct}
In the MCT schematic model, the correlator $\phi(t)$ obeys the
integro-differential equation:
$\phi(t) + t_0 \, \dot \phi(t) + v \int_0^t \mathsf \phi^q(t-s) \ \dot{\phi}(s)\,
  \mathrm{d}s = 0$,
where $q=1$ in the continuous model A and $q=2$ in the discontinuous model B,
$v$ is the controlling parameter and $t_0$ is a characteristic microscopic timescale.
The MCT parameter $\lambda$ for the discontinuous model B is $0.5$.
\bibitem{BBdc} Biroli, G. $\&$ Bouchaud, J.-P.
Critical fluctuations and breakdown of the Stokes-Einstein relation in the mode-coupling theory of glasses,
{\it J. Phys: Condens.} {\bf 19}, 205101 (2007).
\bibitem{het1} 
Cicerone, M. T., Blackburn, F. R. $\&$ Ediger, M. D.
Anomalous diffusion of probe molecules in polystyrene: evidence for spatially heterogeneous segmental dynamics,
{\it Macromolecules} {\bf 28}, 8224 (1995).
\bibitem{het2} 
Kob, W., Donati, C., Plimpton, S. J., Poole, P. H., $\&$ Glotzer, S. C. 
Dynamical heterogeneities in a supercooled Lennard-Jones liquid,
{\it Phys. Rev. Lett.} {\bf 79} 2827 (1997).
\bibitem{het3}
Bennemann, C., Donati, C., Baschnagel, J. $\&$ Glotzer, S. C. 
Growing range of correlated motion in a polymer melt on cooling towards the glass transition,
{\it Nature} {\bf 399}, 246 (1999).
\bibitem{het4}
Franz, S. $\&$  Parisi, G.
On non-linear susceptibility in supercooled liquids,
{\it J. Phys.: Condens. Matter} {\bf 12} 6335 (2000).
\bibitem{het5} 
Widmer-Cooper, A., Harrowell, P.  $\&$  Fynewever, H.
How reproducible are dynamic heterogeneities in a supercooled liquid?
{\it Phys. Rev. Lett.} {\bf  93}, 135701 (2004).
\bibitem{het6} Berthier, L., Biroli, G., Bouchaud, J.-P., Cipelletti, L., El Masri, D., L'Hote, D., Ladieu, F., Pierno, M. 
Direct experimental evidence of a growing length scale accompanying the glass transition,
{\it Science} {\bf 310},1797 (2005).
\bibitem{het7}
Pan, A. C,. Garrahan, J. P., $\&$ Chandler, D.
Heterogeneity and growing length scales in the dynamics of kinetically constrained lattice gases in two dimensions,
{\it Phys. Rev. E} {\bf 72}, 041106 (2005).
\bibitem{het8} 
Bouchaud, J.-P. $\&$ Biroli, G.  
Nonlinear susceptibility in glassy systems: A probe for cooperative dynamical length scales,
{\it Phys. Rev. B} {\bf 72}, 064204 (2005).
\bibitem{het9}  
Biroli, G., Bouchaud, J.-P., Miyazaki, K., $\&$  Reichman,  D. R.
Inhomogeneous mode-coupling theory and growing dynamic length in supercooled liquids,
{\it Phys. Rev. Lett.} {\bf 97}, 195701 (2006).
\bibitem{het10} 
Berthier, L., Biroli, G., Bouchaud, J.-P., Kob, W., Miyazaki, K. $\&$ Reichman, D. R.
Spontaneous and induced dynamic correlations in glass formers. II. Model calculations and comparison to numerical simulations,
{\it J. Chem. Phys.} {\bf 126}, 184503 (2007); J. Chem. Phys. {\bf 126}, 184504 (2007).
\bibitem{het11} 
Chaudhuri, P., Berthier, L. $\&$ Kob, W.
Universal nature of particle displacements close to glass and jamming transitions,
{\it Phys. Rev. Lett.} {\bf 99}, 060604 (2007).
\bibitem{het12} 
Dalle-Ferrier, C., Thibierge, C., Alba-Simionesco, C., Berthier, L., Biroli, G., Bouchaud, J.-P., Ladieu, F., L’Hote, D.
$\&$ Tarjus, G. 
Spatial correlations in the dynamics of glassforming liquids: Experimental determination of their temperature dependence,
{\it Phys. Rev. E} {\bf 76}, 041510 (2007).
\bibitem{conigliodamage}
Coniglio, A., de Arcangelis, L., Herrmann, H. $\&$ Jan, N. Exact relations between damage spreading and thermodynamical properties,
{\it Europhysics Lett.}  {\bf 8}, 315 (1989)
\bibitem{toninelli2}
Toninelli, C., Wyart, M., Berthier, L., Biroli, G., $\&$ Bouchaud J.-P.
Dynamical susceptibility of glass formers: Contrasting the predictions of theoretical scenarios, {\it Phys. Rev. E} {\bf 71}, 041505 (2005).
\bibitem{BB}
Biroli, G.,  $\&$ Bouchaud, J.-P.
Diverging length scale and upper critical dimension in the Mode-Coupling Theory of the glass transition,
{\it Europhys. Lett.} {\bf 67}, 21 (2004).
\bibitem{berthier} 
Berthier, L., Biroli, G., Bouchaud, J.-P., Kob, W., Miyazaki K. $\&$ Reichman, D. R. 
Spontaneous and induced dynamic correlations in glass formers. II. Model calculations and comparison to numerical simulations,
{\it J. Chem. Phys.} {\bf 126}, 184504 (2007).
\bibitem{franz} 
Franz, S., Parisi, G., Ricci Tersenghi, F. $\&$ Rizzo, T.
Field theory of fluctuations in glasses,
{\it Eur. Phys. J. E.} {\bf 34}, 1-17 (2011).


\end{thebibliography}
\end{document}

\title{Supplementary Information
\\
Scaling and universality in glass transition}

\author{Antonio de Candia, Annalisa Fierro, Antonio Coniglio}

\maketitle

The bootstrap percolation (BP) model \cite{CLR} is defined in the following way. First, each site of the lattice is occupied by a particle randomly with 
probability $p$.
Then, each particle that has less than $m$ occupied neighbors ($f$ or more empty neighbors, with $f=k+1-m $, and $k+1$ being the coordination number)
is removed, until a stable configuration is reached. The clusters of particles in this stable configurations are called $m$-clusters. On the Bethe lattice, which we will consider here, 
when $m>2$  the model has a discontinuous transition. In this case there is a probability 
$p_c$,  below which no particles are left at the end of the
procedure, while for $p>p_c$, an infinite $m$-cluster appears with a finite density $P$, that can be considered the order parameter of the model.
The subset of the $m$-cluster, formed by all the sites having exactly $m$ neighbors, is called the ``corona'',
while the sites having more than  $m$ neighbors are called the ``deep core''.

Bootstrap percolation has a mixed order  transition: while the percolation order 
parameter $P$ of BP jumps discontinuously at the threshold from zero to $P_c$, the fluctuation $\chi$ of the order parameter with respect to the initial configuration and the associated length $\xi$ have a critical behavior given by  $P-P_c\sim\epsilon^{\beta}$,
$\chi\sim\epsilon^{-\gamma}$ and $\xi\sim\epsilon^{-\nu}$ where $\epsilon=  |p-p_c|/p_c$
%and the associated length $\xi$ diverge according to:
%\begin{equation}
%P-P_c\sim\epsilon^{\beta},\qquad \chi\sim\epsilon^{-\gamma},\qquad \xi\sim\epsilon^{-\nu},
%\label{eq34}
%\end{equation}

%Although the order parameter is discontinuous, its fluctuation diverge at the critical point,  a behaviour that %characterizes a ``mixed order'' transition.

Here we show that, on the Bethe lattice,
the critical exponent that characterizes the divergence of the fluctuations
of the order parameter is $\gamma=1$ and $\nu=1/4$. For convenience, we also calculate
the critical behavior of the order parameter $P$  with critical exponent  $\beta=1/2$,
the density of sites in the corona, and the divergence of the corona mean cluster size with an exponent $\gamma^\prime=1/2$. 

The Bethe lattice is a lattice extracted randomly from the set of lattices where each site
is connected to $z=k+1$ other sites. Let us consider a Bethe lattice where each site is occupied with probability $p$.
%After the 
%The ``$m$-cluster'' is defined as the set of occupied sites having at least $m$ neighbors belonging to the $m$-cluster.
%The ``corona'' is defined as the set of occupied sites having exactly $m$ neighbors belonging to the $m$-cluster, while the 
%``deep core'' is the $m$-cluster minus the corona, that is the set of occupied sites having more than $m$ neighbors belonging to the $m$-cluster.
Define $\P$ as the probability that a site, that has one of its neighbors belonging to the $m$-cluster, belongs itself to the $m$-cluster.
By recursion, it has to satisfy the self-consistent equation $\P=pF(\P)$, where 
\begin{equation}
F(\P)=\sum\limits_{l=m-1}^{k}\binom{k}{l}\P^l(1-\P)^{k-l}.
\label{eq:1}
\end{equation}
When the occupation probability is greater than a threshold value $p_c$, a solution $\P>0$ appears,
that signals the presence of a $m$-cluster in the system.
For $m\ge 3$, the function $F(\P)$ is a polynomial in $\P$ with lowest order $\P^2$, and therefore the transition cannot be continuous.
When the occupation probability reaches the critical value $p_c$,
the probability $\P$ jumps to a value $\P_c>0$.
Expanding $F(\P)$ in a Taylor series around $\P_c$, and taking into account that $F(\P_c)=\P_c/p_c$ and $F^\prime(\P_c)=1/p_c$, one finds, for $p\to p_c^{+}$,
\[
\P-\P_c\sim |p-p_c|^{1/2}.
\]

\begin{figure}
\begin{center}
\includegraphics[width=6cm]{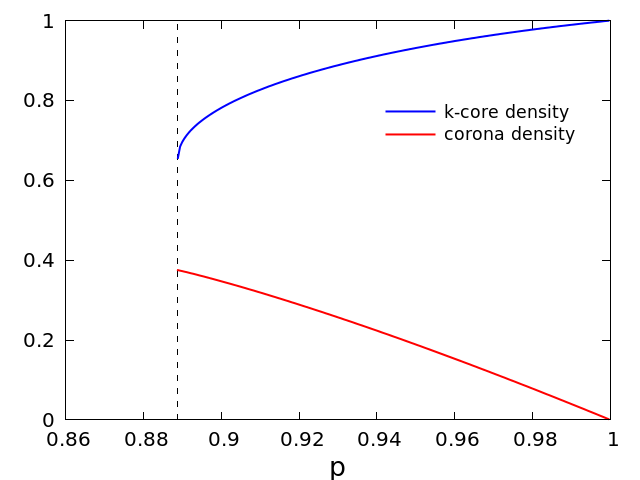}
\caption{Density of the $m$-cluster and of the corona for $k=3$ and $m=3$. The critical point is $p_c=8/9$.}
\label{fig:1}
\end{center}
\end{figure}

The density $P$ of the $m$-cluster and the density $C$ of the corona can be expressed as
\begin{align*}
P&=p\sum\limits_{l=m}^{k+1}\binom{k+1}{l}\P^l(1-\P)^{k+1-l},
\\
C&=p\binom{k+1}{m}\P^m(1-\P)^{k+1-m}.
\end{align*}

These quantities have a singularity at the critical point analogous to the one of $\P$. They jump to a finite value at $p=p_c$, and then they have a $|p-p_c|^{1/2}$
singularity for $p\to p_c^{+}$ (except for the case $k=3$, $m=3$, where the density of the corona is linear near the critical point).
In Fig. \ref{fig:1} we show these densities in the case $k=3$, $m=3$.

Now consider two sites $i$ and $j$ having ``chemical distance'' $n$.
The chemical distance is just the number of steps needed to go from one site to the other, given that on the Bethe lattice in the thermodynamic limit
there is only one path joining the two sites. For a finite size lattice we are neglecting the presence of loops, and therefore multiple paths joining the sites, which is
justified for $n\ll\log(N)$, where $N$ is the total number of sites.
In order to  evaluate the pair correlation function $g_{i,j}$ and the fluctuations of the order parameter $\chi$, 
let us first introduce lattice gas variables  for the  particles of the infinite cluster: 
\[
\pi_i=\left\{\begin{array}{ll}
1&\mbox{if $i$ belongs to the infinite $m$-cluster,}\\
0&\mbox{otherwise.}
\end{array}\right.
\]
It follows that for a lattice of size $N$, 
\begin{subequations}
\begin{align}
P&=\frac{1}{N}\sum_{i=1}^N \langle\pi_i\rangle
\\
g_{i,j}&=\langle\pi_i\pi_j\rangle-\langle\pi_i\rangle\langle\pi_j\rangle
\label{eq3b}
\\
\chi&=\frac{1}{N}\sum_{i,j} g_{i,j}\sim\sum_n k^n g(n)
%\left(\langle\pi_i\pi_j\rangle-\langle\pi_i\rangle\langle\pi_j\rangle\right)\sim\sum_n k^n g(n)
\end{align}
\end{subequations}
where $g(n)=g_{i,j}$ with $i$ and $j$ having chemical distance $n$,
%$g_{i,j}=\langle\pi_i\pi_j\rangle-\langle\pi_i\rangle\langle\pi_j\rangle$,
and
in the last step we took the thermodynamic limit, so that the number of sites at chemical distance $n$ from a given site is proportional to $k^n$.

\begin{figure}
\begin{center}
\includegraphics[width=8cm]{fig2_s.png}
\caption{Possible configurations of the chain of sites from $i$ to $j$, having chemical distance $n$ (see text).}
\label{fig:2}
\end{center}
\end{figure}

We want now to compute $\langle\pi_i\pi_j\rangle$, that is the probability that both $i$ and $j$, having chemical distance $n$,
belong to the infinite $m$-cluster.
In Fig.\ \ref{fig:2} we show all the possible configurations of sites $i$ and $j$, and of the $n-1$ sites between them.
The symbols $p_0$ and $p_1$ represent the possible states of sites $i$ and $j$. $p_0$ ($p_1$) means that the site belongs to the $m$-cluster, has
$m$ or more (exactly $m-1$) neighbors belonging to the $m$-cluster among the $k$ neighbors not shown in figure,
(given that the neighbor shown in figure belongs to the $m$-cluster).  The relative probabilities are
\begin{align*}
P_0&=p\sum_{l=m}^{k}\binom{k}{l}\P^l(1-\P)^{k-l},
\\
P_1&=p\binom{k}{m-1}\P^{m-1}(1-\P)^{k-m+1}.
\end{align*}
On the other hand, symbols $x$, $\theta$ and $\theta_1$  represent the possible states of the $n-1$ sites between $i$ and $j$.
$\theta_1$ ($\theta$) means that the site belongs to the $m$-cluster, has $m-1$ or more (exactly $m-2$) neighbors belonging to the $m$-cluster
among the $k-1$ neighbors not shown in figure, given that at least one (both) the neighbors shown in figure belong to the $m$-cluster.
The relative probabilities are
\begin{align*}
\Theta_1&=p\sum_{l=m-1}^{k-1}\binom{k-1}{l}\P^l(1-\P)^{k-1-l},
\\
\Theta&=p\binom{k-1}{m-2}\P^{m-2}(1-\P)^{k+1-m}.
\end{align*}
Note that $\Theta=\frac{p}{k}F^\prime(\P)$, and therefore at the critical point $\Theta$ jumps to the value $\Theta_c=1/k$. For $p\to p_c^{+}$ it has the same singularity as $\P$, therefore
\[
1-k\Theta\sim |p-p_c|^{1/2}.
\]
Finally, $x$ means that the site can be in any state, and the relative probability is one.

The probability $\langle\pi_i\pi_j\rangle$ is given by the sum of the six probabilities of the configurations of Fig.\ \ref{fig:2}, that are given by
\begin{align*}
P_a&=P_0^2,
\\
P_b&=2P_0P_1\Theta^{n-1},
\\
P_c&=P_1^2\Theta^{n-1},
\\
P_d&=2P_0P_1\Theta_1\sum_{k=0}^{n-2}\Theta^k=2P_0P_1\Theta_1\frac{1-\Theta^{n-1}}{1-\Theta},
\\
P_e&=(n-1)P_1^2\Theta_1\Theta^{n-2},
\\
P_f&=P_1^2\Theta_1^2\sum_{s=0}^{n-3}(s+1)\Theta^k=P_1^2\Theta_1^2\left\{\frac{1-\Theta^{n-1}}{(1-\Theta)^2}-(n-1)\frac{\Theta^{n-2}}{1-\Theta}\right\}.
\end{align*}
The sum of all the terms of $\langle\pi_i\pi_j\rangle$ that do not depend on $n$ gives $\left(P_0+\frac{P_1\Theta_1}{1-\Theta}\right)^2$, that is equal to $P^2$.
This can be verified using the relations
\begin{align*}
P&=P_0+P_1 Q,
\\
Q&=\Theta_1+\Theta Q.
\end{align*}
In the first relation the probability that a site  with $k+1$ neighbors belongs to the $m$-cluster $P$ is given 
by the probability $P_0$ that at least $m$ of the first $k$ neighbors belong to it, plus the probability $P_1 Q$
that $m-1$ among the first $k$ and the $(k+1)$-th neighbor belong to it.
In a similar way the second relation holds for the probability $Q$ that a site belongs to the $m$-cluster under the condition that the first neighbor also belongs to it.      
%The first tells that a site with $k+1$ neighbors belongs to the $m$-cluster if at least $m$ of the first $k$ neighbors belong to it (probability $P_0$),
%or $m-1$ among the first $k$ and the $(k+1)$-th neighbor belong to it (probability $P_1 Q$).
%The second tells the same, with the condition that the first neighbor belongs to the $m$-cluster.
In conclusion, the terms that do not depend on $n$ cancel out with the term $\langle\pi_i\rangle\langle\pi_j\rangle$ in Eq.\ (\ref{eq3b}).
Collecting all the other terms, we have
\[
g(n)=(A+nB)\Theta^n,
\]
where $A$ and $B$ go to a constant finite value for $p\to p_c^{+}$.
As a consequence at leading order 
\[
\chi=\sum_n k^n g(n)\sim \sum_n n(k\Theta)^n\sim \frac{1}{(1-k\Theta)^2}\sim |p-p_c|^{-1},
\]
so that the exponent that describes the fluctuations of the order parameter is $\gamma=1$.

To evaluate the mean cluster size of corona clusters, we have to consider the corona connectivity $\Gamma(n)$,
defined as the probability that two sites at a chemical distance $n$ belong to the same corona cluster.
In this case, the only contributing configurations are the ones of type c) in Fig.\ \ref{fig:2}. Therefore
the connectivity is $\Gamma(n)=P_1^2\Theta^{n-1}$,
and the mean cluster size of corona clusters will be given by
\[
\chi_c=\sum_n k^n \Gamma(n)\sim \sum_n (k\Theta)^n=\frac{1}{1-k\Theta}\sim |p-p_c|^{-1/2},
\]
%where $k^n$ is the number of sites at a chemical distance $n$ from a given site,
so that the critical exponent is $\gamma^\prime=1/2$,
in agreement with Ref.\ \cite{GDM06} and \cite{SLC06}.

\begin{figure}
\begin{center}
\includegraphics[width=6cm]{fig3_s.png}
\caption{Mean cluster size of corona clusters and fluctuations of the order parameter as a function of $p-p_c$.}
\label{fig:3}
\end{center}
\end{figure}

Finally, we consider the exponent $\nu$, that describes how the correlation length $\xi$ diverges approaching the critical point.
It has to be considered that the chemical distance $n$ on the Bethe lattice plays the role of the square of the euclidean distance $r^2$,
because the lattice is infinite dimensional.
We can then define a pair correlation function using  the distance $r$:
\[
\frac{f(r/\xi)}{r^{d-2+\eta}}r^{d-1}dr\sim g(n) k^n dn.
\]
Being $g(n)=n\Theta^n$, we have
\[
\frac{f(r/\xi)}{r^{\eta-1}}dr=r^2(k\Theta)^{r^2}r dr=r^3 f(r/\xi)dr
\]
where
\[
\xi=\sqrt{\left(\ln k\Theta\right)^{-1}}\sim|p-p_c|^{-1/4},
\]
showing that $\nu=1/4$ and $\eta=-2$.
In a similar way for the connectivity of the corona cluster we find the same exponent $\nu = 1/4$ for the correlation length and $\eta=0$, in agreement with the results of Ref.\ \cite{SLC06}.